\documentclass{emulateapj}
\newcommand{\lre}{$\log r_{\rm e}$}
\newcommand{\re}{$r_{\rm e}$}
\newcommand{\mie}{$<\! \mu\! >_{\rm e}$}
\newcommand{\ls}{$\log \sigma_0$}
\newcommand{\dt}{$\delta(\log t)$}
\newcommand{\dz}{$\delta(\log Z)$}
\newcommand{\ct}{$c_{\rm t}$}
\newcommand{\cz}{$c_{\rm Z}$}
\newcommand{\ml}{$ M/L$}
\newcommand{\mls}{$M_{\ast}/L$}

\shorttitle{SDSS-UKIDSS FPs}
\shortauthors{La Barbera et al.}

\begin{document}
\title{The SDSS-UKIDSS Fundamental Plane of Early-type Galaxies}

\author{La Barbera, F. \altaffilmark{1},  
        Busarello, G. \altaffilmark{1},
        Merluzzi, P. \altaffilmark{1},
	de la Rosa, I. \altaffilmark{2},
	Coppola, G. \altaffilmark{3},
	Haines, C.P. \altaffilmark{4}
}

\altaffiltext{1}{INAF -- Osservatorio Astronomico di Capodimonte, Napoli, Italy, }
\altaffiltext{2}{Instituto de Astrofisica de Canarias, Tenerife, Spain,}
\altaffiltext{3}{University of Naples Federico II, Department of Physics, Napoli, Italy,}
\altaffiltext{4}{School of Physics and Astronomy, University of Birmingham,
        Birmingham, UK}

\begin{abstract}
We derive  the Fundamental  Plane (FP) relation  for a sample  of 1430
early-type galaxies in  the optical (r band) and  the near-infrared (K
band),  by  combining  SDSS  and  UKIDSS data.   With  such  a  large,
homogeneous dataset, we are able to assess the dependence of the FP on
the  waveband.   Our  analysis  indicates  that  the  FP  of  luminous
early-type  galaxies  is essentially  waveband  independent, with  its
coefficients increasing at most by  $8\%$ from the optical to the NIR.
This finding fits well into a consistent picture where the tilt of the
FP  is not  driven  by  stellar populations,  but  results from  other
effects, such as non-homology.  In this framework, the optical and NIR
FPs require more massive galaxies  to be slightly more metal rich than
less massive ones,  and to have highly synchronized  ages, with an age
variation per decade in mass smaller than a few percent.
\end{abstract}
\keywords{Galaxies: fundamental parameters -- Galaxies: evolution}

\section{Introduction}
\label{sec:INTR}

Early-type galaxies  (ETGs) populate a two-dimensional  surface in the
space  of parameters  that  reflect size  (effective radius),  density
(mean   surface  brightness),   and   kinetic  temperature   (velocity
dispersion) \citep{George92}.  A key  feature of the Fundamental Plane
(FP;   \citealt{Dressler87};  \citealt{George87})  is   its  deviation
(`tilt')  from the  virial  theorem,  which may  be  interpreted as  a
variation  of the  M/L ratio  along the  sequence of  ETGs  and/or the
breaking  of  homology  assumption,  i.e.   the  fact  that,  for  all
galaxies, the  observed parameters have the  same power-law dependence
on the corresponding physical quantities (namely, the central velocity
dispersion on kinetic energy,  the effective radius on `gravitational'
radius,  and the  effective surface  brightness on  the  overall light
profile,  see e.g.~\citealt{DjS93}).   Despite  all the  observational
efforts, the origin of the tilt  is still under debate.  The change of
the  M/L ratio  can be  explained by  a change  in either  the stellar
population (e.g.~\citealt{PrS96})  or dark matter  content with galaxy
mass  (\citealt{CLR96}). Both  structural  and dynamical  non-homology
have also been  invoked as physical explanations of  the observed tilt
(see            e.g.~\citealt{HjM95},~\citealt{CdC95},~\citealt{GrC97},
and~\citealt{BCC97}).  Recently, ~\citet[from now TBB04]{Tru04} showed
that  the   tilt  is  mostly   driven  by  dynamical   and  structural
non-homology,  while  stellar populations  account  for  only a  small
fraction  of it.~\citet{BBT07}  argued that  the tilt  is  more likely
because of  a variation  of the dark  matter content with  mass, still
favoring a picture where stellar populations play a minor role.  Since
the contribution of different stellar populations to galaxy luminosity
is  expected to  be  wavelength-dependent, while  other effects  (e.g.
non-homology) are not, the dependence of the FP on wavelength directly
informs on how properties of the stellar populations change with mass,
which is a crucial point to understand galaxy formation and evolution.

Previous  studies of  the wavelength  dependence led  to contradictory
results.~\citet{PDdC98b}   and~\citet{SCO98}   found   the   tilt   to
significantly decrease  from optical to  NIR wavelengths, interpreting
this  result  as  an  increase  of  age  and  metallicity  with  mass.
~\citet{MGA99} and~\citet{ZGS02}  found only  a small decrease  of the
tilt with wavelength, with the  FP still being significantly tilted in
the NIR.  However, ~\citet{BER03b},  deriving the FP for ETGs observed
in the Sloan  Digital Sky Survey (SDSS), found evidence  for the FP to
be  wavelength independent  from the  $g$ to  the $z$  bands.  Several
different  effects  can produce  this  puzzling  picture.   The FP  in
different  wavebands has often  been derived  for small  samples, with
inhomogeneous measurements  of galaxy parameters,  different selection
criteria (e.g.  galaxy samples  spanning different ranges in magnitude
and/or velocity dispersion), and  with different fitting methods.  The
FP by~\citet{BER03b} avoided all  these problems by analyzing the same
sample of galaxies  at different wavebands, but it  was limited to the
short wavelength baseline provided by  the SDSS.  In the present work,
for  the first  time,  we derive  the  FP by  using  the same,  large,
homogeneous sample of ETGs  over the wide wavelength baseline provided
by the r- and K-band data of  the SDSS and the UKIRT Infrared Deep Sky
Survey (UKIDSS).

The layout of  the paper is the following. In  Sec.~2, we describe the
selection of the sample, while Sec.~3 details how we obtain the r- and
K-band    structural   parameters,    and    the   central    velocity
dispersions. Sec.~4  deals with  the comparison of  the r-  and K-band
FPs. In  Sec.~5, we  show how  the optical and  NIR FPs  constrain the
variation of stellar population  parameters along the galaxy sequence.
The discussion follows  in Sec.6.  Throughout the paper,  we adopt the
cosmology  $\rm H_0  \!   = \!   75  \, km  \,  s^{-1} \, Mpc^{-1}$,
$\Omega_{\rm m} \!  = \!  0.3$, and $\Omega_{\Lambda} \!  = \!  0.7$.

\section{Sample selection}
\label{sec:DATA}
We select a sample of  ETGs, with available K-band photometry from the
second  data release  of  UKIDSS, and  r-band  photometry and  central
velocity  dispersions  from the  fifth  data  release  (DR5) of  SDSS.
First,  a  complete volume-limited  catalog  of  galaxies is  defined,
consisting of  all the 105036 objects  in DR5 with  an r-band absolute
magnitude M$_{r}{<}-20$, and a  spectroscopic redshift in the range of
0.05 to 0.095.  Absolute magnitudes  are obtained from the SDSS r-band
Petrosian  magnitudes, k-corrected to  redshift $0.1$  by using  the $
kcorrectv4\_1\_4 $ software~\citep{BL03}.  The lower redshift limit is
chosen to minimize the  aperture bias~\citep{GOMEZ03}, while the upper
redshift   limit    guarantees   a   high    level   of   completeness
(see~\citealt{SAR06}).   ETGs  are   defined  according  to  the  SDSS
spectroscopic parameter $eclass$, that classifies the spectral type on
the  basis of  the  principal component  analysis  technique, and  the
photometric  parameter  $fracDev_r$, which  measures  the fraction  of
galaxy  light that  is  fitted  by a  de  Vaucouleurs law.   Following
~\citet{BER03a}, we define as ETGs those objects with $eclass \!  < \!
0$  and $fracDev_r  \!   > \!   0.8$,  resulting in  a  list of  47061
galaxies.   Out of  them, we  retain  only those  33628 galaxies  with
available  central velocity dispersion,  $\sigma_0$, between  $70$ and
$420$  km\,s$^{-1}$.   These  cuts  are required  to  obtain  reliable
$\sigma_0$'s from  the SDSS-DR5  database.  All the  selected galaxies
have spectra with median per-pixel  $S/N$ larger than $10$, which is a
further    requirement   to    retrieve    reliable   SDSS    velocity
dispersions~\footnote{\footnotesize  See the  list of  requirements in
  the    Algorithms   section    of   the    SDSS-DR5    website,   at
  http://www.sdss.org/dr5/algorithms/veldisp.html }.  The SDSS catalog
is cross-matched  with UKIDSS, resulting in 1570  galaxies.  We notice
that all the SDSS galaxies with reliable velocity dispersion, that are
covered by  the UKIDSS  survey, are then  included in this  sample. In
other words,  the matching with  UKIDSS does not change  the magnitude
limit of the present sample. We select only those galaxies observed in
K-band  images with  good seeing  ($FWHM \leq  1''$).   This selection
reduces  the sample size  by only  $10 \%$,  and excludes  cases where
structural parameters might be affected by large uncertainties.  Since
all the r-band images have  FWHM smaller than~$1.4''$ and the galaxies
in our  sample have on average  effective radii larger  in the optical
than in the  K-band (see below), we do not  apply any seeing selection
to the SDSS  photometry.  The above procedure leads  to a final sample
of 1430 galaxies.

\begin{deluxetable*}{c|c|c|c|c|c|c}
 \tablewidth{0pc}
 \tabletypesize{\scriptsize}
 \tablecaption{FP parameters in the $r$ and K bands  for the sample of 1430 gala
 xies.   }
\tablehead{RA & DEC & $\log R_{\rm e,r}$ & $< \!\mu \!>_{\rm e,r}$ &  $\log R_{\rm e,K}$ &  $< \!\mu \!>_{\rm e,K}$ & $\log \sigma_0$ }
 \startdata
   145.34432 &     -0.01692 &   0.860 &   21.602 &   0.527 &   17.046 &   2.282 \\
   147.24805 &     -0.03572 &   0.834 &   21.346 &   0.680 &   17.437 &   2.228 \\
   146.81199 &     -0.19005 &   0.459 &   19.968 &   0.349 &   16.319 &   2.217 \\
   146.09369 &     -0.79309 &   0.985 &   21.521 &   0.475 &   16.216 &   2.259 \\
   146.46892 &     -0.09284 &   1.622 &   23.236 &   1.090 &   18.009 &   2.317 \\
   146.19333 &     -0.03887 &   0.625 &   20.555 &   0.445 &   16.439 &   2.226 \\
   145.68114 &     -0.86722 &   1.001 &   21.576 &   0.721 &   17.350 &   2.196 \\
   145.70894 &     -0.74768 &   0.134 &   18.936 &   0.088 &   15.653 &   2.112 \\
   145.48725 &     -0.80693 &   0.965 &   21.993 &   1.065 &   19.167 &   2.180 \\
   145.42694 &      0.04954 &   0.470 &   19.852 &   0.018 &   14.473 &   2.367 \\
   145.44549 &     -0.12268 &   0.381 &   19.576 &   0.166 &   15.351 &   2.353 \\
   145.34160 &     -0.57727 &   0.087 &   18.124 &  -0.152 &   13.574 &   2.355 \\
   145.19382 &      0.16887 &   0.192 &   18.862 &   0.137 &   15.669 &   1.930 \\
   146.28017 &     -0.40695 &   0.033 &   17.760 &  -0.129 &   13.996 &   2.288 \\
   147.30829 &      0.15116 &   0.830 &   21.046 &   0.454 &   16.295 &   2.287 \\
   146.72794 &     -0.55688 &   0.262 &   19.223 &   0.178 &   15.409 &   2.316 \\
   148.85664 &     -0.05916 &   0.978 &   21.643 &   1.160 &   19.055 &   2.226 \\
   147.79347 &      0.12326 &   0.614 &   20.337 &   0.227 &   15.566 &   2.187 \\
   147.74868 &      0.11584 &   0.480 &   20.362 &   0.125 &   15.682 &   2.233 \\
   148.58499 &     -0.94207 &   1.180 &   22.575 &   0.703 &   17.637 &   2.109 \\
   149.12382 &     -0.39828 &   0.365 &   19.682 &  -0.037 &   14.726 &   2.324 \\
   149.11298 &     -0.34883 &   0.685 &   20.638 &   0.593 &   16.954 &   2.270 \\
   149.18631 &     -0.31181 &   1.291 &   23.217 &   0.823 &   18.064 &   2.101 \\
   148.84251 &     -0.04411 &   0.469 &   20.031 &   0.387 &   16.276 &   2.202 \\
   149.11264 &     -0.47563 &   0.572 &   20.065 &   0.471 &   16.069 &   2.398 \\
   149.17153 &     -0.41298 &   0.359 &   19.539 &   0.266 &   15.894 &   2.260 \\
 \enddata
 \end{deluxetable*}

\section{FP parameters}
\label{sec:FP}
The  photometric parameters  entering  the FP,  namely, the  effective
radius,  $r_{\rm e}$,  and  the mean  surface  brightness within  that
radius,  \mie, are  derived using  2DPHOT~\citep{LBdC08}.  The  r- and
K-band  images are  processed  by adopting  the  same 2DPHOT  options,
allowing homogeneous structural parameters  to be derived between both
bands.   For each galaxy,  a local  PSF model  is computed  by fitting
simultaneously the  four closest  stars to that  galaxy on  the image.
Structural parameters,  i.e.  the effective parameters  and the Sersic
index $n$ (shape  parameter), are then derived by  modeling the galaxy
images with two-dimensional seeing-convolved Sersic models.  Effective
radii are  converted to physical  units by using the  angular diameter
distance corresponding to the DR5 spectroscopic redshift, $z$, of each
galaxy.    Mean  surface  brightnesses   are  de-reddened   using  the
extinction  maps  of~\citet{SFD98},  are  corrected  for  cosmological
dimming,  by  subtracting the  term  $\rm  7.5  \log (1+z)$,  and  are
k-corrected  to   redshift  $0.1$   with  the  $   kcorrectv4\_1\_4  $
software~\citep{BL03}.    In  order  to   estimate  the   accuracy  on
structural parameters,  we use 160,  out of the 1430,  galaxies having
repeated observations  in different UKIDSS  frames.  We find  that the
averaged  differences  between  repeated  measurements  of  structural
parameters are fully consistent with zero, amounting to $-0.01 \!  \pm
\!  0.01$, $-0.037 \!  \pm \!   0.04$, and $0.005 \!  \pm \! 0.01$ for
$\log r_{\rm e}$, \mie \,  and $\log n$, respectively.  The rms values
of  these  differences  amount  to   $32  \%$  in  \re,  $0.6  \rm  \,
mag/arcsec^2$ in \mie, and $25 \%$ in $n$.  Notice that the scatter in
$\log r_{\rm e}$ is fully  consistent with the typical accuracy of the
measurement  of the  half-light radii~\citep{Kelson00}.   The quantity
$\log r_{\rm e}  \!  - \!  0.3 \!   < \!  \mu \!  >_{\rm  e}$, that is
the relevant photometric parameter entering the FP, has an uncertainty
of  only $7 \%$,  as expected  due to  the correlation  of measurement
errors of the  effective parameters.  The comparison of  r- and K-band
structural parameters for the  present sample is fully consistent with
what  is found  in  our previous  studies (e.g.~\citealt{LMB04}).   In
particular, the  mean ratio between r-  and K-band radii  is $-0.11 \!
\pm \!  0.01$dex,  i.e.  on average effective radii  decrease by $\sim
\!  25 \%$ from the optical to the NIR.  This value is consistent with
that of  $\sim 20  \%$ found by~\citet{PdCD98a},  and is  in agreement
with the existence of negative color gradients in early-type galaxies.
Sersic  indices are  fully consistent  between optical  and  NIR.  The
average ratio of r- to K-band  $n$ values amounts to $-0.007 \! \pm \!
0.009$dex.
Central velocity dispersions are  retrieved from DR5 and are corrected
as  in  \citet{BER03b}  to  a  relative  aperture  of  $r_{\rm  e}/8$,
following~\citet{JFK95}.  As shown by~\citet{BERN07}, for $\sigma_0 \!
<  \! 150$  km  s$^{-1}$  the DR5  velocity  dispersions are  slightly
overestimated.  This small  bias increases up to $12  \%$ at $\sigma_0
\sim  100$  km  s$^{-1}$.   We  remove this  effect  by  applying  the
correction curve shown in  fig.~4 of~\citet{BERN07} (see the grey line
in the upper-left  panel)\footnote{\footnotesize The SDSS-DR6 velocity
  dispersions are  not affected by  this bias, but they  are available
  only  for   $85\%$  of  our  sample.   However,   we  verified  that
  restricting the analysis to the sample with DR6 velocity dispersions
  changes the FP coefficients by less than $2\%$.}.  The r- and K-band
effective  parameters,   as  well   as  the  corrected   DR5  velocity
dispersions are  given in Tab.~1 (fully available  in electronic form)
for all the  1430 galaxies analyzed in the  present study.  Columns in
the table  provide the  following quantities.  Columns~1  and~2 report
right  ascension  (RA) and  declination  (DEC)  in  units of  degrees.
Columns~3 and~4 provide the logarithm of the effective radius (in unis
of  $kpc$) and  the  effective  mean surface  brightness  (in unis  of
$mag/arcsec^2$)  in  the  r-band.    Columns~5  and~6  list  the  same
quantities as  columns~3 and~4 but  for the K-band.   Column~7 reports
the corrected DR5 velocity dispersions.

\begin{figure*}[t!]
\epsscale{0.9}
\plotone{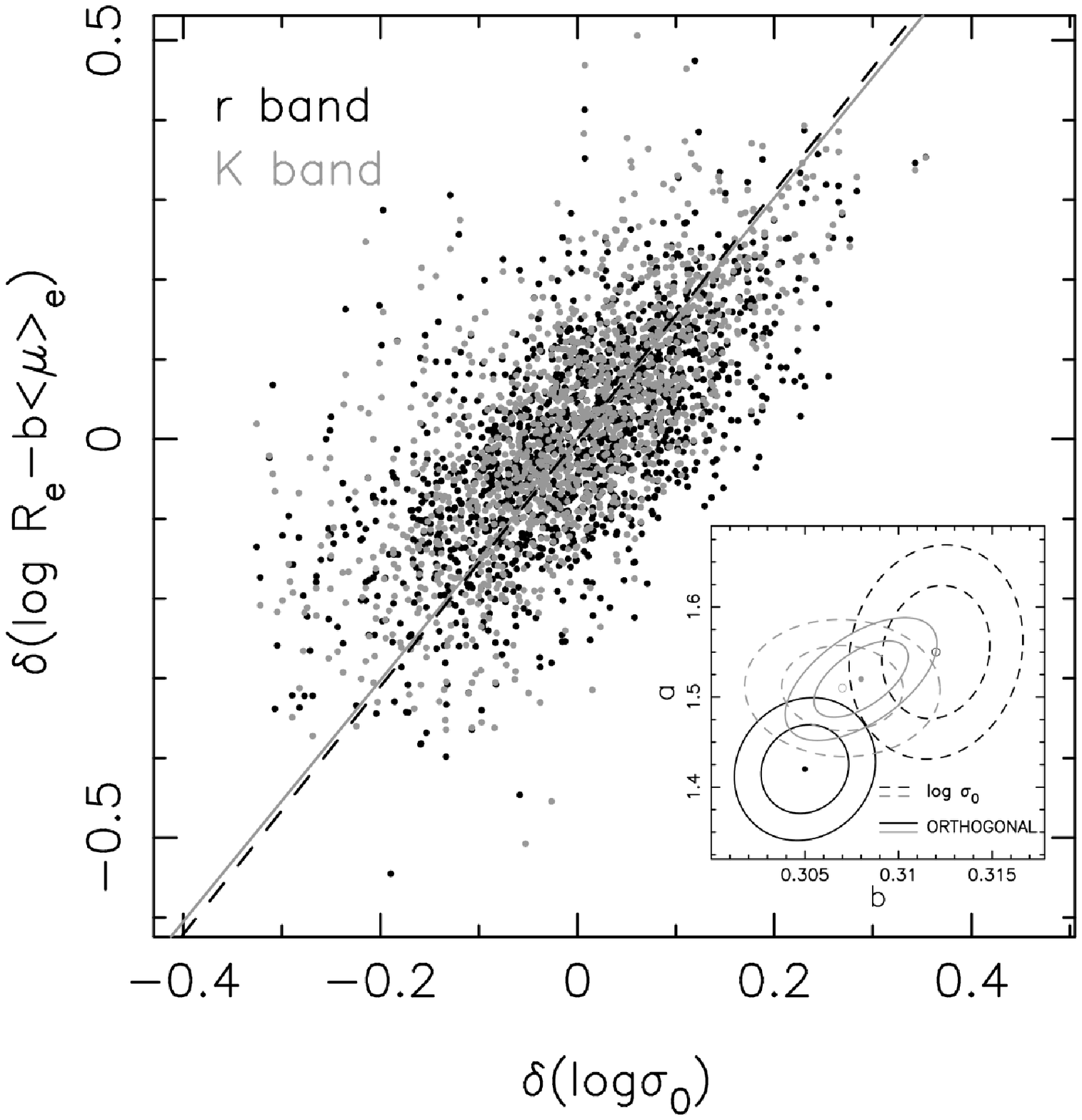}
\caption{Short  edge-on  view of  the  r-  and  K-band FPs,  with  the
  photometric quantity, $\log R_{\rm e} \!  - \!  b < \!  \mu \!>_{\rm
    e}$, being plotted as a function of \ls. Both quantities have been
  normalized  by subtracting  their median  values, enabling  a direct
  comparison of the r- and K-band projections. The median \ls \, value
  amounts to $2.215 \rm dex$ .   The FPs are represented by two lines,
  whose  slopes  are equal  to  the coefficients  $a$  of  the \ls  \,
  fits.  Black  and grey  colors  correspond to  the  r  and K  bands,
  respectively. The  average value of  $b=0.308$ was adopted  for both
  bands.  The inset compares the r-  and K-band slopes of the FP, with
  the  concentric  ellipses denoting  $1$  and  $2 \sigma$  confidence
  contours.  Results of the orthogonal and \ls \, fits are represented
  by solid and dashed ellipses, respectively.}
\label{fig:FP_EDGE}
\end{figure*}

\section{The SDSS and UKIDSS Fundamental Planes}
\label{sec:RES}

We  adopt  the usual  representation  of the  FP,  $\log  r_{\rm e}  =
a \log \sigma_{\rm 0}  + b < \!\mu\! >_{\rm e} + c  \ $, where $a$ and
$b$ are  the ``slopes'',  and $c$ is  the offset.   These coefficients
were derived  by minimizing the  sum of the absolute  residuals around
the plane.   This method has the  advantage of being  less affected by
outliers (e.g.~\citealt{JFK96}).   We adopted two  fitting procedures,
minimizing  the  residuals  either  in  \ls\,  or  in  the  orthogonal
direction to the plane.  The  first method is virtually independent of
selection effects  in the plane  of photometric parameters,  while the
orthogonal  fit,  adopted  in  most  previous works,  treats  all  the
variables symmetrically (see~\citealt{LAB00}).

The  FP  coefficients were  corrected  for  selection effects  through
Monte-Carlo  simulations.   First,   we  generated  galaxy  magnitudes
according  to the r-band  luminosity function  of ETGs~\citep{BER03a}.
For  each  magnitude,  we  derived~\lre   \,  and  \mie  \,  from  the
luminosity--size  relation of~\citet{Shen03}.  Values  of \ls  \, were
assigned by  using the  equation of the  FP, assuming given  values of
$a$, $b$, and  $c$, and a given scatter in  \ls.  All these quantities
were chosen  to match the observed  FP. Notice that  when deriving the
simulated FP  we applied the same  cuts in magnitude and~\ls  \, as we
did for  the real  sample.  The corrections  for selection  effects on
$a$,  $b$,  and  $c$, were  estimated  by  not  applying the  cuts  in
magnitude and~\ls \, to the simulated FP.  These corrections amount to
$+0.01\%$, $+8\%$ and $+5\%$ for the  \ls \, fit, and $+35 \%$, $+7\%$
and    $+17\%$    for     the    orthogonal    fit.     As    expected
(e.g.~\citealt{LAB00}),   the   magnitude   cut   underestimates   the
coefficient $a$  of the orthogonal fit,  while for the \ls  \, fit the
effect  is negligible.   The above  corrections depend  mainly  on the
scatter around the FP, and, because of the very similar dispersions of
the r- and K-band FPs (see below), were applied to both the r- and the
K-band coefficients.  We notice  that the above procedure assumes that
our  sample of  early-type galaxies  is magnitude  complete.  However,
because of  the matching of  the initial volume-complete  SDSS catalog
with the UKIDSS  database, that reduces the sample  size from 33628 to
1430 ETGs, the above assumption  might not necessarly hold. To address
this point, we retrieved effective parameters and velocity dispersions
for the whole sample of 33628  ETGs in the SDSS catalog, and estimated
how  the r-band  FP  coefficients  change between  the  whole and  the
UKIDSS-matched  samples.   We found  the  variation  to be  completely
negligible, amounting  to $1 \%$ and to  $ 2\%$ for the  values of $a$
and $b$ obtained by the orthogonal fit.

\begin{deluxetable}{c|c|c|c|c}
\tablecolumns{5}
\tablewidth{0pc}
\tablecaption{FP coefficients.~\label{tab:FPCOF}}
\tablehead{& a & b & c & $\rm rms$ }
\startdata
$\log \sigma_o $ fit &&&&\\
$\rm r \! - \! band$ & $1.55 \pm 0.07$ & $0.312 \pm 0.003$ & $-9.1 \pm 0.1$ & $0.081 $ \\
$\rm K \! - \! band$ & $1.51 \pm 0.04$ & $0.307 \pm 0.003$ & $-8.6 \pm 0.1$ & $0.073 $ \\
\hline
orthogonal fit &&&&\\
$\rm r \! - \! band$ & $1.42 \pm 0.05$ & $0.305 \pm 0.003$ & $-8.8 \pm 0.1$ & $0.064 $ \\
$\rm K \! - \! band$ & $1.53 \pm 0.04$ & $0.308 \pm 0.003$ & $-8.6 \pm 0.1$ & $0.062 $ 
\enddata
\end{deluxetable}

The coefficients  of the FP are reported  in Tab.~\ref{tab:FPCOF}.  In
Fig.~\ref{fig:FP_EDGE},  we compare the  edge-on views  of the  r- and
K-band planes,  and in the  inset we show  the values of $a$  and $b$.
The $a$ coefficients differ by  only $2\sigma$ for the orthogonal fit,
and  they almost  coincide  for the  \ls  \, fit,  which is  virtually
unaffected by selection  effects.  The value of $b$  ($\sim 0.308$) is
independent of the  waveband, as well as the scatter  of the FP, which
presents a  tiny difference  ($<2 \%$)  for the \ls  \, fit  (see also
~\citealt{PDdC98b}).  The r-band value  of $a$ is fully consistent with
that of $a=1.49 \pm 0.05$  found by~\citet{BER03b}, while it is larger
than that of $a=1.24 \pm 0.07$ found by~\citet{JFK96}. For the K-band,
the \ls  \, coefficient  is consistent with  the value of  $a=1.53 \pm
0.08$  found  by~\citet{PDdC98b},  while  it  is larger  than  that  of
$a=1.38 \pm  0.1$ found by~\citet{ZGS02}.  We notice  that, because of
the large sample  size, the accuracy of our  K-band FP coefficients is
significantly  higher  (by  $50\%$)  than in  previous  studies.   The
invariance of the FP with waveband is in agreement with~\citet{Cap06},
who  found for  25  ETGs from  the  SAURON project  the  M/L versus  L
relation to have the same slope in both the I- and K-bands.

\begin{figure*}[t!]
\epsscale{0.9}
\plotone{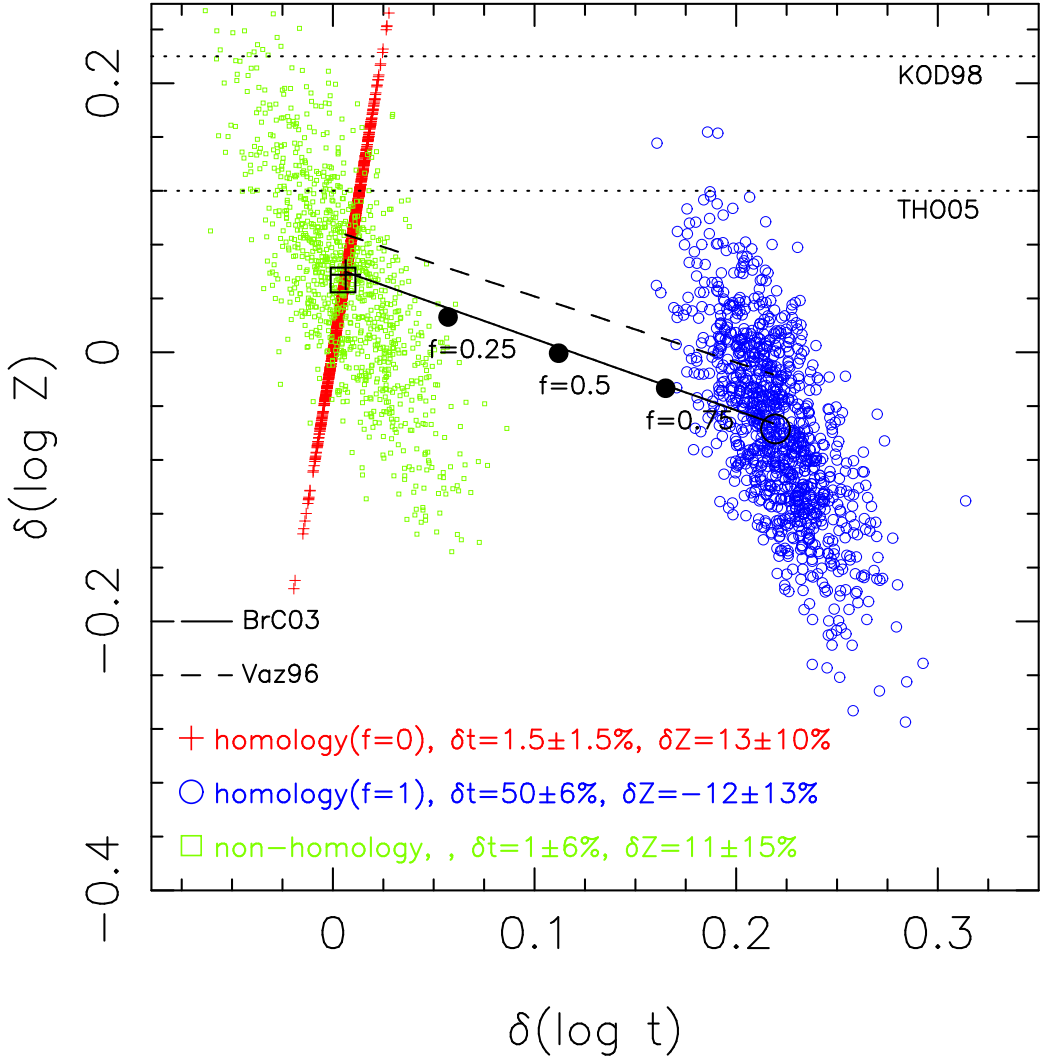}
\caption{Differences of age, \dt, and metallicty, \dz, per decade 
in mass between more and less massive galaxies, as derived from the FP
slopes in the r and K bands, by adopting the mean values of $a$ of the
orthogonal  and \ls  \,  fits (see  the  text). Red  crosses and  blue
circles correspond to the cases  of homology assumption with $f=0$ and
$f=1$, respectively,  while the green  squares correspond to  the case
where non-homology is taken into account (see Sec.~5). For each symbol
type, different points  mark the values of \dt \,  and \dz \, obtained
by shifting $a_{\rm r}$ and $a_{\rm K}$ according to the corresponding
uncertainties. For  each case, the  median values, $\delta \rm  t$ and
$\delta \rm Z$, of \dt \, and  \dz \, are plotted as black symbols and
are  reported in  the  lower-left part  of  the plot.   The effect  of
changing $f$ on the median values of \dt \, and \dz \, is shown by the
black  solid and dashed  lines, for  the BrC03  and the  Vaz96 models,
respectively.   The  horizontal  dotted   lines  mark  the  values  of
$\delta \rm  Z$ obtained from previous studies  of the color-magnitude
relation (KOD98) and absorption line indices (THO05).}
\label{fig:DLOGT_DLOGZ}
\end{figure*}

\section{Constraints on the stellar populations}
\label{sec:ANALYSIS}

The  tilt of  the  FP can  be  parameterized as  a power-law  relation
between \ml \, and $M$. We assume that the stellar mass-to-light ratio
of galaxies, \mls, is  a power-law: \mls$\propto M^{\beta^\ast}$. This
agrees with  what was found in  previous studies for  bright ETGs (see
TBB04).  The  \ml \, vs.  $M$  relation can then be  written as $M/L\!
\propto  \!  M^{\beta  +\beta^\ast}$ (see  also~\citealt{PrS96}).  The
quantity $\beta$ is  related to the ratio of stellar  to total mass as
$M_{\ast}/M\!   \propto \!  M^{-\beta}$,  describing the  variation of
wavelength  independent  properties   with  mass  (e.g.   dark  matter
content).  The quantity $\beta^\ast$ depends on how stellar population
parameters vary  with mass.  Considering  only the effects of  age and
metallicity, for each waveband, we can write:
\begin{eqnarray}
\frac{\delta(\log \! M_{\ast}/L)}{\delta(\log \! M_{\ast})} =\frac{\beta^{\ast}}{1-\beta}=c_{\rm t} \cdot \frac{ \delta(\log \! t) }{\delta(\log \! M_{\ast})} + c_{\rm Z} \cdot \frac{\delta(\log \! Z)}{\delta(\log \! M_{\ast})},
\label{eq:mlchange}
\end{eqnarray}
where the quantities \dt \, and \dz \, are the logarithmic differences
of age and metallicity defined  between more and less massive galaxies
(per decade in stellar  mass), while $c_{\rm t}=\frac{\partial \log \!
  M_{\ast}/L}{\partial \log t}$  and $c_{\rm Z}=\frac{\partial \log \!
  M_{\ast}/L}{\partial \log  Z}$ are the partial  derivatives of $\log
\!   M_{\ast}/L$  (in a  given  band) with  respect  to  $t$ and  $Z$.
Writing  Eq.~\ref{eq:mlchange}  for  r  and  K bands,  we  obtain  two
independent  constraints on  \dt  \, and  \dz.   Then, expressing  the
values of $\beta^\ast$ in the  r and K bands ($\beta^\ast_{\rm r}$ and
$\beta^\ast_{\rm K}$) as a  function of the corresponding coefficients
of the FP, we can estimate  \dt \, and \dz. We introduce the parameter
$f=\beta^\ast_{\rm K}  / (\beta  + \beta^\ast_{\rm K})$  which defines
the  fraction of  the K-band  slope of  the \ml  \, vs.   $M$ relation
(i.e.  the fraction  of the  K-band  tilt) due  to stellar  population
effects.We note that $f$ can vary between $0$ and $1$.  For $f=0$, the
K-band tilt is  independent of stellar populations ($\beta^{\ast}_{\rm
  K}=0$), while  for $f=1$ the  tilt is entirely explained  by stellar
population effects ($\beta=0$).  Under the assumption of homology, the
slope  of  the  \ml  \,  vs.   $M$  \,  relation  can  be  written  as
$(2-a)/(2+a)$.  With the above notation, the following relations hold:
\begin{eqnarray}
 \beta+\beta^\ast_{\rm r}  = (1-f)/f \! \cdot \! \beta^\ast_{\rm K}+\beta^\ast_{\rm r}     =  (2-a_{\rm r})/(2+a_{\rm r}), \nonumber & (2a) \\
  \beta+\beta^\ast_{\rm K} =  \beta^\ast_{\rm K}/f =  (2-a_{\rm K})/(2+a_{\rm K}). \nonumber & (2b) 
\label{eq:mlslopes}
\end{eqnarray}
For   different    $f$'s   and   using   the   values    of   $a$   in
Tab.~\ref{tab:FPCOF},    we   computed   $\beta^\ast_{\rm    r}$   and
$\beta^\ast_{\rm K}$ from  Eqs.~\ref{eq:mlslopes}, and then, inverting
Eq.~\ref{eq:mlchange} for both bands, we  derived \dt \, and \dz.  The
values  of  \ct \,  and  \cz \  were  estimated  using simple  stellar
population   models,   with   solar   metallicity  and   an   age   of
$12$~Gyr~\footnote{Computing \ct \, and  \cz \, by varying $t$ between
$5$ and $12$Gyr,  and $Z$ from $0.5 Z_{\odot}$  to $2.0 Z_{\odot}$, we
found that the  change of the median  values of \dt \, and  \dz \, are
negligible,   amounting  to  less   than  $0.01$dex   and  $0.005$dex,
respectively.}, using both the  \citet{BrC03} (from now BrC03) and the
updated  \citet{Vazdekis:96} (from  now Vaz96)  models.  We  adopted a
Scalo  IMF  and  a  Salpeter  IMF  for the  BrC03  and  Vaz96  models,
respectively.  Fig.~\ref{fig:DLOGT_DLOGZ}  shows \dz \,  versus \dt \,
obtained for  $f=0$ and $f=1$,  as well as  the mean values of  \dt \,
and  \dz \,  as  a function  of $f$.   The  scatter seen  in the  plot
reflects  the  uncertainties listed  in  Tab.~\ref{tab:FPCOF} for  the
$a_{\rm r}$ and $a_{\rm K}$ coefficients. The figure shows that if the
NIR  tilt  of the  FP  is not  caused  by  stellar population  effects
($f=0$), more  massive galaxies have to  be more metal  rich than less
massive ones (\dz$>0$), with  galaxy ages being remarkably homogeneous
($\delta t/t \sim 1 \%$).  As $f$ increases, we see that
\dz \, decreases, while \dt \, becomes larger. Specifically,  for 
$f   \sim  1$,  more   massive  galaxies   are  much   older  ($\delta
t/t \sim50 \%$) and less metal rich than low mass systems.

\section{Discussion}
\label{sec:DISC}

This work presents the waveband  dependence of the FP by comparing the
optical and NIR  FPs for a large sample  of galaxies, with homogeneous
measurements of structural  parameters and velocity dispersions.  This
is allowed,  for the  first time, thanks  to the availability  of both
SDSS r-band photometry and  spectroscopy, and UKIDSS K-band photometry
for the same sample of ETGs.  Such a dataset, together with the use of
the same  fitting procedure in  both bands, makes our  study virtually
free from any methodological effect  on the waveband dependence of the
FP.  Our analysis shows that the FP does not change significantly from
the  optical to  the  NIR, bringing  interesting  questions about  the
nature of the sequence of ETGs.

In  Sec.~5, we  have shown  how the  r- and  K-band FPs  constrain the
variation of  the stellar population properties  (age and metallicity)
with stellar mass, and how  such a constraint is strongly dependent on
the fraction,  $f$, of the  FP tilt resulting from  stellar population
effects.  Previous studies of the color-magnitude (CM) relation and of
line-strength indices of ETGs might  help us to solve this dependency,
by deriving  the proper value  for $f$. \citet{Kod98} showed  that the
little  redshift evolution  of the  CM  relation implies  (i) all  the
(luminous) ETGs to be equally old and (ii) more massive galaxies to be
more metal  rich, with a metallicity  change of $\delta  (\log Z) \sim
0.22$dex per  decade in stellar  mass.  This finding  is qualitatively
consistent  with  that  of~\citet{THO05},  who  found  absorption-line
indices consistent with a metallicity  change of $\delta (\log Z) \sim
0.12$dex  per  mass decade.   They  also  found  evidence for  an  age
gradient along the sequence of ETGs, with $\delta (\log t) = +0.05 \pm
0.07$  (see their  eq.3).   Fig.~\ref{fig:DLOGT_DLOGZ} compares  these
values of $\delta (\log Z)$ with those derived by the optical--NIR FP.
For the BrC03  (Vaz96) model, the maximum value  of $\delta (\log Z)$,
which is consistent with the  FP, amounts to $0.06 $($0.09$) $\pm 0.04
$($0.04$)dex for $f=0$. This value  is $4 \sigma$ ($3.7 \sigma$) lower
than that  derived by the CM  relation, while it is  only $1.5 \sigma$
($0.8  \sigma$)  lower  than  that  found  by~\citet{THO05}.   As  $f$
increases, the FP requires the value of $\delta (\log Z)$ to decrease,
making  the  above differences  even  larger.  Therefore,  reconciling
previous estimates  of $\delta (\log Z)$  with our results  leads to a
scenario where  $f=0$, which means that  the FP tilt is  not driven by
stellar  populations.    We  have   to  remark,  however,   that  this
interpretation is  troublesome, since  galaxy colors and  line indices
are always measured within a given fixed aperture, and the presence of
internal population gradients in galaxies can significantly affect the
inferred  values   of  $\delta  (\log   Z)$  and  $\delta   (\log  t)$
(e.g.~\citealt{Scodeggio:01}).

Further constraints come from  previous works addressing the origin of
the  FP tilt  itself.   Performing a  detailed  dynamical analysis  of
twenty-five galaxies,  \citet{Cap06} derived \ml  \, ratios consistent
with  those obtained  from the  virial  theorem in  the assumption  of
homology, concluding  that structural and orbital  non-homology have a
negligible role in the tilt  of the FP (see also~\citealt{ZZG08}).  In
support  to this  view, they  also showed  that the  variation  of the
dynamical \ml \, is  correlated with the H$_\beta$ line-strength, thus
ascribing   most   of   the   tilt   to   stellar   population   (age)
effects. However, as the  authors notice, this result strictly applies
to their measurement of the velocity dispersion as the average over an
aperture of radius equal to  $r_e$, a fact that alone might compensate
part of the dynamical non-homology.  Moreover, most of the galaxies in
their sample (68\%) are fast  rotators, while five of them (20\%) have
low  velocity  dispersion  ($\sigma$=60-85  km  s$^{-1}$).   As  found
by~\citet{ZGZ06}  and~\citet{DOF08}, bright  and faint  spheroids have
different FPs, with  the tilt becoming larger for  galaxies having low
velocity dispersion.  Hence, the different selection of our sample and
that   of~\citet{Cap06}   prevents   a   straightforward   comparison.
\citet{BBT07} showed  that by  replacing mean surface  brightness with
mass density,  the FP relation  closely approaches the  virial theorem
expectation, implying that  most of the tilt is  caused by a variation
of dark  matter content with galaxy mass.   However, the uncertainties
on their  FP coefficients  and the possible  biases introduced  by the
gravitational-lens   selection  do   not   definitively  exclude   the
contribution of non-homology  to the FP tilt.  Our  result agrees with
\citet{BBT07} regarding  the minor role played  by stellar populations
on  the tilt.   TBB04,  agreeing with~\cite{BCC97}  and~\citet{GrC97},
found that structural and  dynamical non-homology can account for more
than  two-thirds  of  the  FP  tilt, with  the  remaining  part  being
explained by  stellar population effects.   In particular, restricting
the analysis to the  magnitude-complete subsample, they found that the
contribution of  stellar populations  to the tilt  becomes negligible.
Notice also  that their FPs  are derived from different  sources, with
significantly different coefficients in the optical and NIR wavebands.

To understand  how our  results may be  affected by the  assumption of
homology,  we followed  an approach  similar to  that of  TBB04, using
spherical,  isotropic,  non  rotating,  one-component models  of  ETGs
following the  Sersic law (see~\citealt{LBC05}).  For  each galaxy, we
considered  the  model with  the  corresponding  Sersic  index in  the
r-band, and used that model to correct the central velocity dispersion
to the  quantity $\sigma_{\rm t}$  (defined as the square-root  of the
total  specific   kinetic  energy),   the  effective  radius   to  the
gravitational radius,  $r_{\rm g}$, and to calculate  the mean surface
brightness within the gravitational radius, $ <\!  \mu \!  >_{\rm g}$.
Applying the orthogonal  fit, we obtain the following  equation of the
FP in the K band: $\log  r_{\rm g} \propto (2.3 \pm 0.2) \log \sigma_t
+ (0.4 \pm 0.02) <\!  \mu \!  >_{\rm g}$, which is remarkably close to
the virial theorem expectation, implying that non-homology may account
for the entire tilt.  To  explore how this result would affect stellar
population properties, we normalized the r- and K-band FP coefficients
in such  a way  to match the  virial theorem expectation  ($a=2.0$ and
$b=0.4$) in the K band.   Then, we derived the corresponding values of
\dz \,  and \dt  \, (see Sec.~\ref{sec:ANALYSIS}).   As we can  see in
Fig.~\ref{fig:DLOGT_DLOGZ}, accounting  for non-homology leads  to the
same \dz \, and \dt \, values as those derived under the assumption of
homology if the tilt of the FP is not due to stellar populations (i.e.
$f=0$).   As discussed above,  this is  consistent with  what expected
from the color-magnitude relation and absorption-line indices of ETGs.


In summary, our  analysis suggests a consistent picture  where (i) the
tilt of the FP does  not originate from stellar population effects but
is due  to other effects,  such as non-homology; (ii)  the SDSS-UKIDSS
FPs require  more massive galaxies to  be mildly more  metal rich than
less massive  systems, and to  have extremely synchronized  ages, with
the age variation per mass decade being smaller than few percent.  


%

\acknowledgements

We  thank  R.R.   de  Carvalho, S.G.   Djorgovski,  M.Capaccioli,  and
A.Mercurio  for   the  helpful  comments  and   suggestions.  We  also
acknowledge the referee for his/her helpful suggestions. We also thank
A.   Vazdekis for providing  us with  the most  recent version  of his
stellar population code.  We have  used data from the 2nd data release
of the UKIDSS survey (\citet{Law07}),  which is described in detail in
\citet{War07}.  Funding for the SDSS  and SDSS-II has been provided by
the Alfred  P.  Sloan Foundation, the  Participating Institutions, the
National  Science  Foundation, the  U.S.   Department  of Energy,  the
National   Aeronautics   and   Space  Administration,   the   Japanese
Monbukagakusho,  the  Max Planck  Society,  and  the Higher  Education
Funding Council for England.

\end{document}